\def\be{\begin{equation}}
\def\ee{\end{equation}}
\def\bea{\begin{eqnarray}}
\def\eea{\end{eqnarray}}

\documentclass[12pt]{article}
\usepackage{epsfig,latexsym}
\begin{document}
\thispagestyle{empty} \vspace*{0.5 cm}
\begin{center}
{\large \bf Renormalization-Group flow for the field strength in
scalar self-interacting theories}
\\
\vspace*{1cm} {\bf M. Consoli and D. Zappal\`a  }\\ \vspace*{.3cm}
{\it INFN, Sezione di Catania}\\
{\it Via S. Sofia 64, I-95123, Catania, Italy} \\
\vspace*{.6cm}

\vspace*{1 cm}
{\bf ABSTRACT} \\
\end{center}

We consider the Renormalization-Group coupled equations for the
effective potential $V(\phi)$ and the field strength $Z(\phi)$ in
the spontaneously broken phase as a function of the infrared cutoff
momentum $k$. In the $k \to 0$ limit, the numerical solution of the
coupled equations, while consistent with the expected convexity
property of $V(\phi)$, indicates a sharp peaking of $Z(\phi)$ close
to the end points of the flatness region that define the physical
realization of the broken phase. This might represent further
evidence in favor of the non-trivial  vacuum field renormalization
effect already discovered with variational methods.
\\
\vskip 0.5 cm
\noindent
Pacs 11.10.Hi , 11.10.Kk

\parskip 0.3cm
\vspace*{3cm}
\vfill\eject
\setcounter{page}{1}
\voffset -1in
\vskip2.0cm

\newcommand{\fa}{\phi^a}
\newcommand{\fb}{\phi^b}
\newcommand{\p}{\partial_{\mu}}
\newcommand{\dd}{\delta^{ab}}
\newcommand{\nn}{\nonumber}


{\bf 1.}~ Renormalization Group (RG) techniques originally inspired
to the Kadanoff-Wilson blocking procedure \cite{kad}  represent a
powerful method to approach non-perturbative phenomena in quantum
field theory. A widely accepted technique consists in starting from
a bare action defined at some ultraviolet cutoff $\Lambda$ and
effectively integrating out shells of quantum modes down to an
infrared cutoff $k$. This procedure provides a $k-$dependent
effective action $\Gamma_k[\Phi]$ that evolves into the full
effective action $\Gamma[\Phi]$ in the limit $k\to 0$.

The $k-$dependence of $\Gamma_k[\Phi]$ is determined by a
differential functional flow equation that is known in the
literature in slightly different forms
\cite{weg,nicol,polch,wet1,mor1}. In particular, with the flows
discussed in detail in Ref.\cite{wetreport} one starts form first
principles and obtains a class of functionals that interpolates
between the classical bare Euclidean action and the full effective
action of the theory. However, some features, such as the basic
convexity property of the effective action for $k \to 0$
\cite{tetradis,litim2,alexander,zappala}, are independent of the
particular approach.

In this Letter, we shall study the coupled equations for the
$k-$dependent effective potential $V_k(\phi)$ and field strength
$Z_k(\phi)$, which naturally appear in a derivative expansion of
$\Gamma_k[\Phi]$ (around the coordinate independent configuration  $\Phi(x)=\phi$ ), 
using a proper-time infrared regulator. This
approach generates evolution equations which are well defined
truncations of the first-principle flows within a background field
formulation \cite{litim12,litim11}. Incidentally, as shown in \cite{litim12},
they are also an approxmation to another exact flow 
(a generalised Callan-Symanzik flow) without  
background fields.

The equations for $V_k(\phi)$
and $Z_k(\phi)$, derived in Ref.\cite{bonanno}, correspond to such
first-principle flows up to additional terms explicitly determined
in Ref.\cite{litim12}. An indication of the reliability of such an
approximation is provided by the consistent computation of the
critical exponents in scalar self-interacting theories for various
numbers $D$ of the space-time dimensions and $N$ of field components
\cite{bonanno}. At the same time, in Ref.\cite{zappala}, it was
shown that going beyond the approximation $Z=1$ is essential to
reproduce successfully the energy gap between the exact ground state
and the first excited state of the double well potential in the
quantum-mechanical limit of the theory $D=1$. In addition, the
coupled equations obtained by using the proper time regulator are
not affected by singularities and/or ambiguities that instead appear
using a sharp infrared cutoff \cite{io2l}.

With these premises, assuming the possibility to neglect the
additional terms of Ref.\cite{litim12} and adopting the same
notations of Refs.\cite{zappala,bonanno}, we shall start our
analysis considering the two equations \be \label{vdim} k{{\partial
V}\over{\partial k}} = \left({k^2\over{4\pi}}\right)^{D/2}
e^{-V''/(Z k^2)} \ee \bea &&k{{\partial Z}\over{\partial k}} = \left({k^2\over{4\pi}}\right)^{D/2}  e^{-V''/(Z k^2)}\nonumber\\
&&\times\Biggl( -{{Z''}\over{Z k^2}}+{{(4+18D-D^2)
(Z')^2}\over{24Z^2k^2}} +{{(10-D)Z'V'''}\over{6 (Z k^2)^2}}-
{{Z(V''')^2}\over{6(Z k^2)^3}}\Biggr ) \label{zdim} \eea where we
have set $V=V_k(\phi)$, $Z=Z_k(\phi)$ and used the notation
$V',Z'$,...to indicate differentiation with respect to $\phi$. In
the following, we shall first analyze these two equations to
understand the approach to convexity  and obtain informations on $Z$
and finally provide a possible physical interpretation of our
numerical results. \vskip 15 pt

{\bf 2.}~~For a numerical solution of Eqs.(\ref{vdim}) and
(\ref{zdim}) it is convenient to use dimensionless  variables
defined as follows: $t={\rm ln}(\Lambda/k)$, $x=k^{1-D/2}\phi$,
$V(t,x)=k^{-D} V_k(\phi)$ and $Z(t,x)=Z_k(\phi)$. By defining the
first derivative of the effective potential $f(x,t)=\partial_x
V(t,x)$, Eqs.(\ref{vdim},\ref{zdim}) become
 \be \label{fadim}
{{\partial f}\over{\partial t}} ={{(D+2)}\over {2}}f + {{(2-D)}\over
{2}}~x{{\partial f}\over{\partial x}} - {{1}\over{(4\pi)^{D/2}}}~
{{\partial }\over{\partial x}} e^{-f'/Z}  \ee \bea \label{zadim}
{{\partial Z}\over{\partial t}}&=& {{(2-D)}\over {2}}~x{{\partial
Z}\over{\partial x}}+
 {{1}\over{(4\pi)^{D/2}}}~
{{\partial }\over{\partial x}} \left (
{{Z'}\over{Z}}e^{-f'/Z}\right)\nonumber\\
-{{e^{-f'/Z}}\over{(4\pi)^{D/2} }}~
 &\Biggl(&{{f'(Z')^2}\over{Z^3}}+{{18D-D^2-20}\over{24}}
{{(Z')^2}\over{Z^2}} +{{(4-D)Z'f''}\over{6 Z^2}}- {{(f'')^2}\over{6
Z^2}}\Biggr ) \eea It is easy to show that these two coupled
equations can be transformed into the structure ($i,j$=1-3) \be
P_{ij} {{\partial U_j}\over{\partial t}} + Q_i= {{\partial
R_i}\over{\partial x}} \ee where the components of the vector
$U_i(x,t)$ are the unknown functions of the problem and where
$P_{ij}$, $Q_i$ and $R_i$ can depend on $x,t, U_i, {{\partial
U_i}\over{\partial x}}$.

In this way, the numerical solution has been obtained with the help
of the NAG routines that integrate a system of non-linear parabolic
partial differential equations in the $(x,t)$ two-dimensional plane. The
spatial discretisation is performed using a Chebyshev $C^o$
collocation method, and the method of lines is employed to reduce
the problem to a system of ordinary differential equations. The
routines contain two main parameters (the size of the discretisation
grid $N_{\rm points}$ in the spatial dimension and the local
accuracy $\Delta$ in the time integration) that can be changed to
control the stability of the solution. In our case, after reaching sufficiently
large values of $N_{\rm points}$ ($\sim 2000$) and sufficiently
small values of $\Delta$ ($\sim 10^{-7}$), the numerical results
remain remarkably stable for further variations of
these parameters.

We shall focus on the quantum-field theoretical case $D=4$ assuming
standard boundary conditions at the cutoff scale : i) a
renormalizable form for the bare, broken-phase potential \be
\label{bare} V_\Lambda(\phi)=-{{1}\over{2}}M^2\phi^2 + \lambda
\phi^4 \ee and ii) a unit renormalization condition for the
derivative term in the bare action \be Z_\Lambda(\phi)=1 \ee With
this choice of the classical potential the problem is manifestly
invariant under the interchange $\phi \to -\phi$ at all values of
$k$.

We shall also concentrate on the weak coupling limit $\lambda=0.1$,
fixing $M=1$ and $\Lambda=10$. In this way, one gets a well defined
hierarchy of scales where the infrared region corresponds to the
limit $k \ll M\ll \Lambda$.

Before addressing the full problem, we have considered the standard
approximation of setting $Z=1$. In this case, we can find a good
approximation to the exact solution of Eq.(\ref{fadim}) for large
$x$ and large $t$ as a cubic polynomial with $t-$dependent
coefficients \be \label{asy}
  f_{\rm asy}(x,t) =A(t)x^3+B(t)x^2+C(t)x +D(t) \ee
This type of form, motivated by RG arguments, see e.g.
\cite{litim3}, becomes exact where $f'$ is large and positive so
that $e^{-f'}\to 0$ yielding $A(t)=A_0$, $B(t)=B_0\exp(t)$,
$C(t)=C_0\exp(2t)$, $D(t)=D_0\exp(3t)$. Higher powers $x^n$, with
$n>3$, might also be inserted but they are suppressed by exponential
terms $e^{-(n-3)t}$. At the same time, $f=0$ is also a solution of
Eq.(\ref{fadim}) and corresponds to a flat effective potential in
$x$. As we shall see, the $k-$evolution, for $k\to 0$, turns out to
be a sort of way to interpolate between $f=0$ and the asymptotic
trend in Eq.(\ref{asy}).

We show in Fig.1 the plots of $V'_k(\phi)\equiv {{\partial
V_k(\phi)}\over{\partial \phi}}$ at various values of the infrared
cutoff down to the smallest value $k=0.05$ that can be numerically
handled by our integration routine. This corresponds to a maximal
value $t_{\rm max}\sim 5.3$ which is comparable to the highest  value
$t=5$ reported in Fig.1a of Ref.\cite{litim2}. In our problem, this
seems to mark the boundary of the region that is very difficult to
handle numerically.

As one can see, the approach to convexity is very clean, in full
agreement with the general theoretical arguments of
Ref.\cite{litim2}. It is characterized by an almost linear behaviour
in the inner $\phi-$region that matches with the outer, asymptotic
cubic shape discussed above. In this sense, the limitation in the maximal $t$
is just a numerical
artifact of our integration method since from the physical point of
view there should be no conceptual problem in the limit $t \to
\infty$.  This is supported by a fit to the slope $S(k)$
of $V'_k(\phi)$ at $\phi=0$  in Fig.1,  which suggests the following  functional form
($a$ and $b$ being  numerical coefficients obtained from the fit)
\be
S(k)\sim  a [{\rm exp}(-bk^2) -1]
\ee
that vanishes in the limit $k \to 0$.

For small $k$, the minimum of the first
derivative of the effective potential $\phi=\hat{\phi}(k)$ does not
correspond to an analytic behaviour. This is a well known result of
quantum field theory \cite{syma}: the Legendre-transformed effective
potential is not an infinitely differentiable function.

Let us now consider the full problem defined by Eqs.(\ref{fadim})
and (\ref{zadim}). Again, for large $x$ and $t$, the pair $f=f_{\rm
asy}(x,t)$ and $Z=1$ provide a simultaneous solution. However, for
finite values of $k$, where the potential is not yet convex downward
and $f'<0$, the term $e^{{{|f'|}\over{Z}}}$ drives $Z$ to grow.

We show in Fig.2, the simultaneous solutions for $Z_k(\phi)$ (upper
panel) and  $V'_k(\phi)\equiv {{\partial V_k(\phi)}\over{\partial
\phi}}$ (lower panel) for various values of the infrared cutoff down
to $k=0.07$ (again  $t_{\rm max}\sim 5$)
that represents, for the coupled problem, the point
beyond which the integration routines no longer work. 
The strong peaking in
$Z_k(\phi)$, to a very high accuracy, occurs at $\hat{\phi}(k)$
where $V'_k(\phi)$ has its minimum. On the basis of the general
convexification property this point tends, for $k \to 0$, to the end
point $\hat{\phi}(0)\equiv \phi_0$ of the flatness region that
defines the physical realization of the broken phase. Finally, we
report in Fig.3 the values of $Z_k(\phi)$ at $\phi=0$ (circles) and
at the peak for $\phi=\hat{\phi}(k)$ (diamonds) in the range
$0.07\leq k \leq 0.15$. Using some extrapolation forms, the value of
the peak seems to tend, in the limit $k \to 0$, to a very large but
finite value $Z_{k=0}(\phi_0)$.

\vskip 15 pt {\bf 3.}~Let us now explore a possible physical
interpretation of the numerical results reported above.
We start by observing that spontaneous symmetry breaking is usually
considered a semi-classical phenomenon, i.e. based on a classical
potential with perturbative quantum corrections. These, with our
choice of the bare parameters $\lambda=0.1$ and $M=1$ in
Eq.(\ref{bare}), and our cutoff value $\Lambda=10$, are typically
small for all quantities. In particular $Z$, in perturbation theory,
is a non-leading quantity since its one-loop correction is
ultraviolet finite. Therefore, the deviations from unity are
expected to be very small.

Re-writing the $\Phi^4$ term in the standard form ${{\lambda_{\rm
st}}\over{4!}}\Phi^4$, with $\lambda_{\rm st}=2.4$, the perturbative
prediction is  \be \label{zpert} Z_{\rm pert}-1={\cal O}(
{{\lambda_{\rm st}}\over{16\pi^2}})\sim 10^{-2} \ee This is also
consistent with the assumed exact "triviality" property of the
theory for $D=4$ \cite{book} that requires $Z \to 1$ in the
continuum limit $\Lambda \to \infty$.

Now, let us compare this prediction with our $Z_k(\phi)$ in Fig.2.
For large values of the infrared cutoff, when the effective
potential is still smooth, we find $Z_k(\phi)\sim 1$ for all values
of $\phi$, as expected. However, at smaller $k$, say $k <\delta$
with $\delta\sim 0.15$, there are large deviations from unity in the
region of $\phi$ where the smooth form of the perturbative potential
evolves into the typical non-analytical behaviour of the exact
effective potential. This leads to the observed strong peaking
phenomenon at the point $\hat{\phi}(k)$, where $V'_k(\phi)$ reaches
its minimum value. This point, on the basis of the general
convexification property of $V(\phi)$ tends, for $k \to 0$, to the
value $\hat{\phi}(0)=\phi_0$, the end point  of the flatness region
that defines the physical realization of the broken phase.

If we express the full scalar field $\Phi(x)$ as \be
\Phi(x)=\phi+h(x) \ee the above results indicate that the higher
frequency components of the fluctuation field $h(x)$, those with
4-momentum $p_\mu$ such that $\delta \leq |p|\leq \Lambda$,
represent genuine quantum corrections for all values of the
background field $\phi$ in agreement with their perturbative
representation as weakly coupled massive states.

On the other hand, the components with a 4-momentum $p_\mu$ such
that $ |p| \leq \delta$,  are non-perturbative for values of the
background field in the range $0 \leq \phi \leq \hat{\phi}(|p|)$. In
particular, the very low-frequency modes with $|p| \to 0$ behave
non-perturbatively for all values of the background in the full
range $0 \leq \phi \leq \phi_0$. They can be thought as collective
excitations of the scalar condensate and cannot be represented as
standard massive states.

The existence of a peculiar $p_\mu \to 0$ limit in the broken phase,
for which one can give some general arguments \cite{consoli}, finds
support in the results of lattice simulations of the broken-symmetry
phase (see Ref.\cite{cea}). There, differently from what happens in
the symmetric phase, the connected scalar propagator deviates
significantly from (the lattice version of) the massive
single-particle form $1/(p^2+{\rm const})$ for $p_\mu \to 0$. In
particular, looking at Figs. 7, 8 and 9 of Ref.\cite{stevenson}, one
can clearly see that, approaching the continuum limit of the lattice
theory, these deviations become more and more pronounced but also
confined to a smaller and smaller region of momenta near $p_\mu=0$.

This observation suggests that the existence of a non-perturbative
infrared sector in a region $0\leq |p| \leq \delta$ might not be in
contradiction with the assumed exact "triviality" property of the
theory if, in the continuum limit, the infrared scale $\delta$
vanishes in units of the physical parameter $m$ associated with the
massive part of the spectrum. This means to establish a hierarchy of
scales $\delta \ll m \ll \Lambda$ such that ${{\delta}\over{m}}\to
0$ when ${{m}\over{\Lambda}} \to 0$.

If this happens, the region $0\leq |p| \leq \delta$ would just
shrink to the zero-measure set $p_\mu=0$, for the continuum theory
where $m$ sets the unit mass scale, thus recovering the exact
Lorentz covariance of the energy spectrum since the point $p_\mu=0$
forms a Lorentz-invariant subset. In this limit, the RG function
$Z_k(\phi)$ would become a step function which is unity for all
finite values of $k$ (and $\phi$) and is only singular for $k=0$ in
the range $0 \leq \phi \leq \phi_0$. In this way, one is left with a
massive, free-field theory for all non-zero values of the momentum,
and the only remnant of the non-trivial infrared sector is the
singular re-scaling of $\phi$ (the projection of the full scalar
field $\Phi(x)$  onto $p_\mu=0$).

This is precisely the scenario of Refs.\cite{alternative}, where for
$\Lambda \to \infty$ the re-scaling of the scalar condensate
diverges logarithmically as $\sim \ln\Lambda$ and the re-scaling of
the finite-momentum modes tends to unity. The existence of such a
divergent re-scaling factor for the vacuum field would have
potentially important phenomenological implications for the scalar
sector of the standard model and for the validity of the generally
accepted upper bounds on the Higgs boson mass \cite{alternative}.

Of course, for a more precise comparison with the numerical results
obtained in this Letter, one should study the value of the peak in
$Z_k(\phi)$ at different values of the bare parameters to check the
predicted logarithmic behaviour $Z_{\rm peak}\sim \ln (\Lambda)$
that, at the present, is just a conjecture suggested by previous
works. In turn, this requires to improve on the present integration
routines to extend the solution of the RG equations towards the
point $k=0$ thus reducing the arbitrariness associated with
different extrapolation forms. 

At the same time, the existence of a
peak in $Z_k(\phi)$ is an interesting feature that would deserve to
go beyond the present numerical analysis.
In this respect, since, according to \cite{laca}, $V_k(\phi)$, 
for $k\to 0$,  is only  
weakly dependent  on the type of infrared regulator, we expect  
 the peak  of Z to be a physical phenomenon and not
an artifact  of the regulator. 
This can be checked by re-considering the
problem from scratch at the level of the exact flow equations.

\vfill\eject

\begin{figure}
\psfig{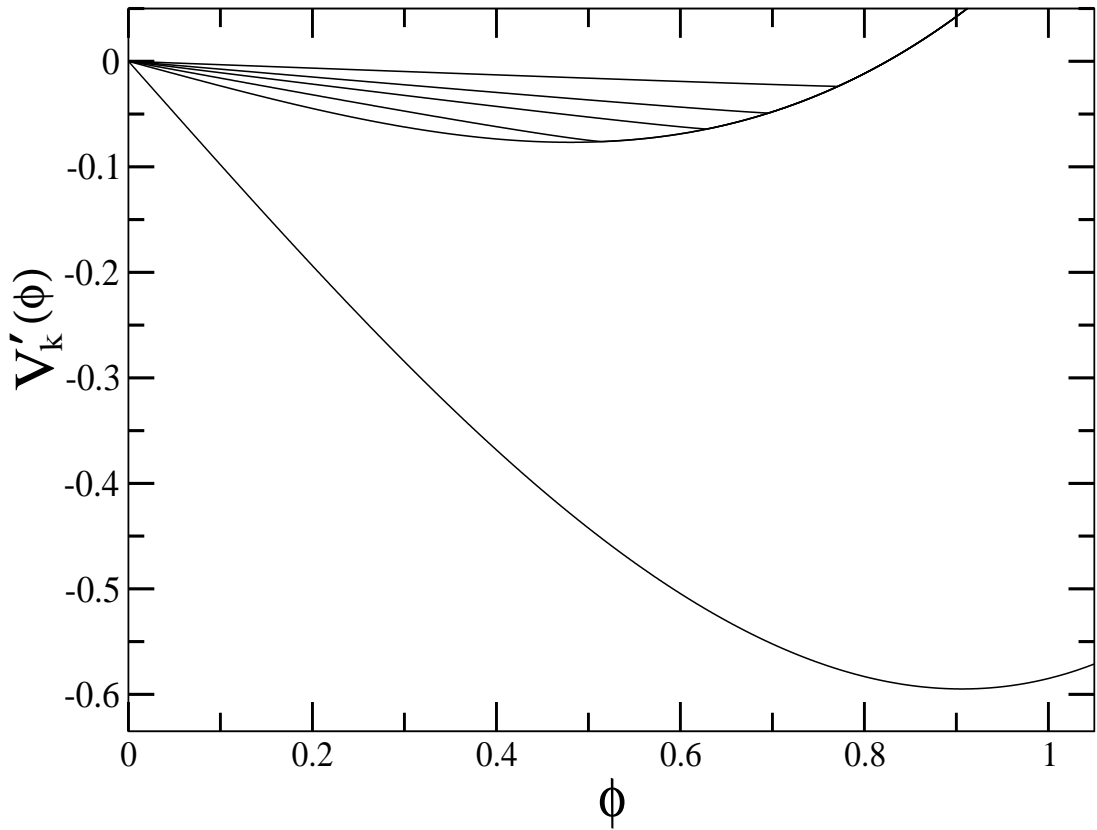} \caption{ ${V'}_k(\phi)$  {\it vs.} $\phi$, with $Z=1$
fixed, at  various values of the infrared cutoff $k$: the lowest
curve is for $k=\Lambda=10$, and then, from bottom to the top,
$k=0.3,~0.13,~ 0.1,~0.08,~0.05 $. }
\end{figure}

\vfill\eject

\begin{figure}
\psfig{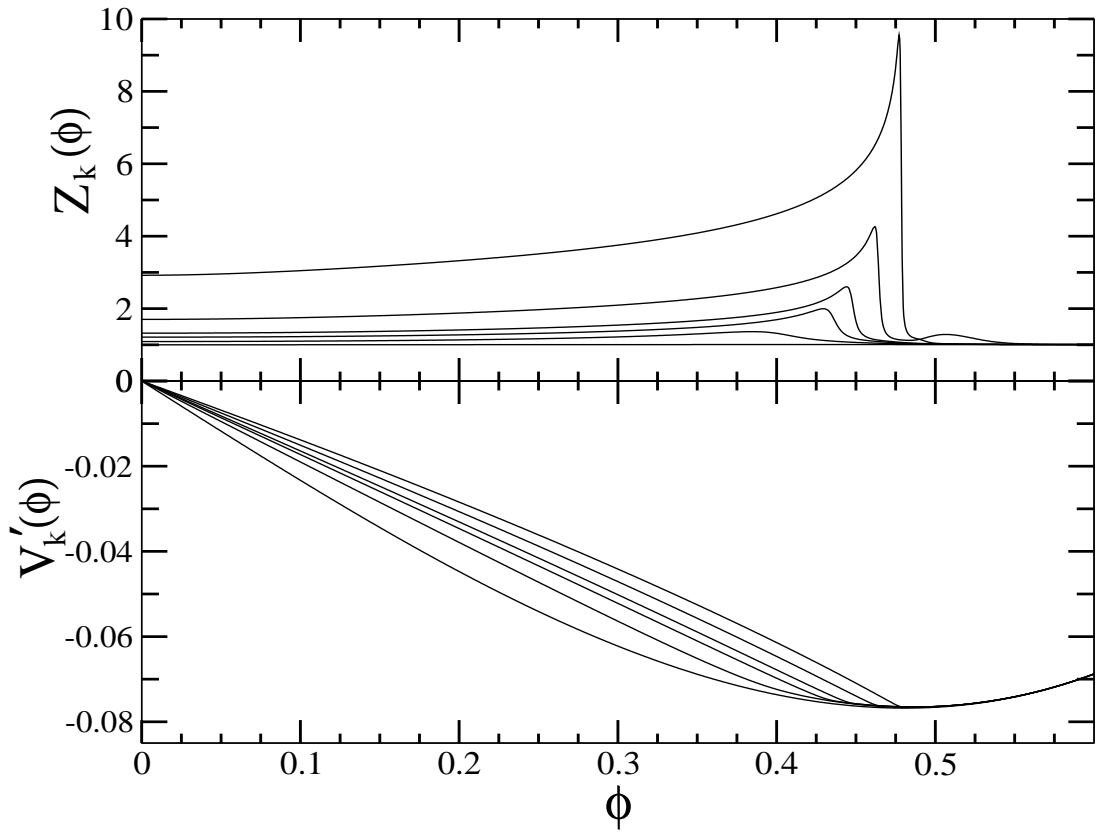} \caption{   $Z_k(\phi)$ (upper panel) and  $
{V'}_k(\phi)$ (lower panel), {\it vs.} $\phi$
 at  $k=$0.3, 0.15, 0.13, 0.12, 0.1, 0.07, from bottom to the top.
}
\end{figure}

\begin{figure}
\psfig{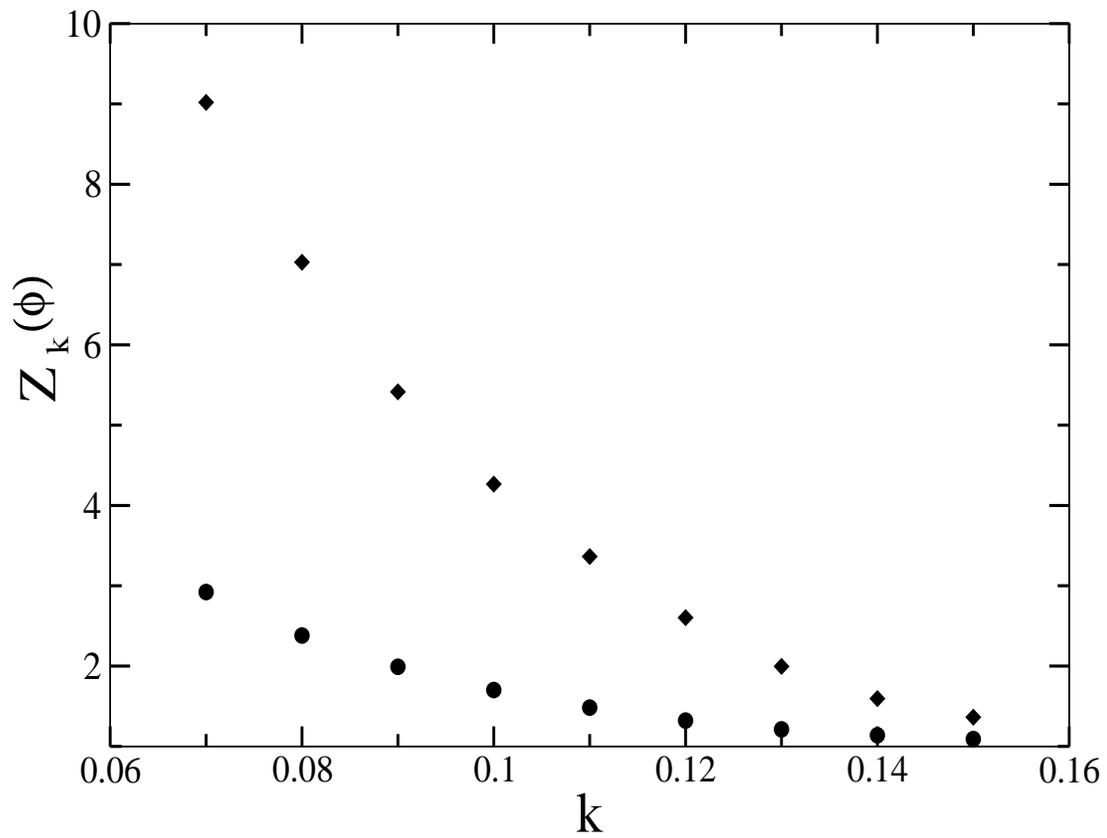} \caption{$Z_k(\phi)$ at the origin $\phi=0$ (circles)
and at the peak  $\phi=\hat{\phi}$ (diamonds), at different values
of the infrared cutoff $k$. }
\end{figure}
\end{document}